\documentclass[aps,prl,twocolumn,showpacs,groupedaddress]{revtex4}

\usepackage{graphicx}
\usepackage{dcolumn}
\usepackage{bm}      
\usepackage{amssymb}

\begin{document}

\hspace{5.2in} \mbox{FERMILAB-PUB-06/085-E}

\title{Measurement of the $B^{0}_{s}$ Lifetime Using Semileptonic Decays}


\author{                                                                      
V.M.~Abazov,$^{36}$                                                           
B.~Abbott,$^{76}$                                                             
M.~Abolins,$^{66}$                                                            
B.S.~Acharya,$^{29}$                                                          
M.~Adams,$^{52}$                                                              
T.~Adams,$^{50}$                                                              
M.~Agelou,$^{18}$                                                             
J.-L.~Agram,$^{19}$                                                           
S.H.~Ahn,$^{31}$                                                              
M.~Ahsan,$^{60}$
% ...
\\                                                                            
\vskip 0.30cm                                                                 
\centerline{(D\O\ Collaboration)}                                             
\vskip 0.30cm                                                                 
}                                                                             
\affiliation{                                                                 
\centerline{$^{1}$Universidad de Buenos Aires, Buenos Aires, Argentina}       
\centerline{$^{2}$LAFEX, Centro Brasileiro de Pesquisas F{\'\i}sicas,         
                  Rio de Janeiro, Brazil}                                     
\centerline{$^{3}$Universidade do Estado do Rio de Janeiro,                   
                  Rio de Janeiro, Brazil}                                     
\centerline{$^{4}$Instituto de F\'{\i}sica Te\'orica, Universidade            
                  Estadual Paulista, S\~ao Paulo, Brazil}                     
\centerline{$^{5}$University of Alberta, Edmonton, Alberta, Canada,           
                  Simon Fraser University, Burnaby, British Columbia, Canada,}
\centerline{York University, Toronto, Ontario, Canada, and                    
                  McGill University, Montreal, Quebec, Canada}                
\centerline{$^{6}$Institute of High Energy Physics, Beijing,                  
                  People's Republic of China}                                 
\centerline{$^{7}$University of Science and Technology of China, Hefei,       
                  People's Republic of China}                                 
\centerline{$^{8}$Universidad de los Andes, Bogot\'{a}, Colombia}             
\centerline{$^{9}$Center for Particle Physics, Charles University,            
                  Prague, Czech Republic}                                     
\centerline{$^{10}$Czech Technical University, Prague, Czech Republic}        
\centerline{$^{11}$Center for Particle Physics, Institute of Physics,         
                   Academy of Sciences of the Czech Republic,                 
                   Prague, Czech Republic}                                    
\centerline{$^{12}$Universidad San Francisco de Quito, Quito, Ecuador}        
\centerline{$^{13}$Laboratoire de Physique Corpusculaire, IN2P3-CNRS,         
                   Universit\'e Blaise Pascal, Clermont-Ferrand, France}      
\centerline{$^{14}$Laboratoire de Physique Subatomique et de Cosmologie,      
                   IN2P3-CNRS, Universite de Grenoble 1, Grenoble, France}    
\centerline{$^{15}$CPPM, IN2P3-CNRS, Universit\'e de la M\'editerran\'ee,     
                   Marseille, France}                                         
\centerline{$^{16}$IN2P3-CNRS, Laboratoire de l'Acc\'el\'erateur              
                   Lin\'eaire, Orsay, France}                                 
\centerline{$^{17}$LPNHE, IN2P3-CNRS, Universit\'es Paris VI and VII,         
                   Paris, France}                                             
\centerline{$^{18}$DAPNIA/Service de Physique des Particules, CEA, Saclay,    
                   France}                                                    
\centerline{$^{19}$IReS, IN2P3-CNRS, Universit\'e Louis Pasteur, Strasbourg,  
                    France, and Universit\'e de Haute Alsace,                 
                    Mulhouse, France}                                         
\centerline{$^{20}$Institut de Physique Nucl\'eaire de Lyon, IN2P3-CNRS,      
                   Universit\'e Claude Bernard, Villeurbanne, France}         
\centerline{$^{21}$III. Physikalisches Institut A, RWTH Aachen,               
                   Aachen, Germany}                                           
\centerline{$^{22}$Physikalisches Institut, Universit{\"a}t Bonn,             
                   Bonn, Germany}                                             
\centerline{$^{23}$Physikalisches Institut, Universit{\"a}t Freiburg,         
                   Freiburg, Germany}                                         
\centerline{$^{24}$Institut f{\"u}r Physik, Universit{\"a}t Mainz,            
                   Mainz, Germany}                                            
\centerline{$^{25}$Ludwig-Maximilians-Universit{\"a}t M{\"u}nchen,            
                   M{\"u}nchen, Germany}                                      
\centerline{$^{26}$Fachbereich Physik, University of Wuppertal,               
                   Wuppertal, Germany}                                        
\centerline{$^{27}$Panjab University, Chandigarh, India}                      
\centerline{$^{28}$Delhi University, Delhi, India}                            
\centerline{$^{29}$Tata Institute of Fundamental Research, Mumbai, India}     
\centerline{$^{30}$University College Dublin, Dublin, Ireland}                
\centerline{$^{31}$Korea Detector Laboratory, Korea University,               
                   Seoul, Korea}                                              
\centerline{$^{32}$SungKyunKwan University, Suwon, Korea}                     
\centerline{$^{33}$CINVESTAV, Mexico City, Mexico}                            
\centerline{$^{34}$FOM-Institute NIKHEF and University of                     
                   Amsterdam/NIKHEF, Amsterdam, The Netherlands}              
\centerline{$^{35}$Radboud University Nijmegen/NIKHEF, Nijmegen, The          
                  Netherlands}                                                
\centerline{$^{36}$Joint Institute for Nuclear Research, Dubna, Russia}       
\centerline{$^{37}$Institute for Theoretical and Experimental Physics,        
                   Moscow, Russia}                                            
\centerline{$^{38}$Moscow State University, Moscow, Russia}                   
\centerline{$^{39}$Institute for High Energy Physics, Protvino, Russia}       
\centerline{$^{40}$Petersburg Nuclear Physics Institute,                      
                   St. Petersburg, Russia}                                    
\centerline{$^{41}$Lund University, Lund, Sweden, Royal Institute of          
                   Technology and Stockholm University, Stockholm,            
                   Sweden, and}                                               
\centerline{Uppsala University, Uppsala, Sweden}                              
\centerline{$^{42}$Physik Institut der Universit{\"a}t Z{\"u}rich,            
                   Z{\"u}rich, Switzerland}                                   
\centerline{$^{43}$Lancaster University, Lancaster, United Kingdom}           
\centerline{$^{44}$Imperial College, London, United Kingdom}                  
\centerline{$^{45}$University of Manchester, Manchester, United Kingdom}      
\centerline{$^{46}$University of Arizona, Tucson, Arizona 85721, USA}         
\centerline{$^{47}$Lawrence Berkeley National Laboratory and University of    
                   California, Berkeley, California 94720, USA}               
\centerline{$^{48}$California State University, Fresno, California 93740, USA}
\centerline{$^{49}$University of California, Riverside, California 92521, USA}
\centerline{$^{50}$Florida State University, Tallahassee, Florida 32306, USA} 
\centerline{$^{51}$Fermi National Accelerator Laboratory,                     
            Batavia, Illinois 60510, USA}                                     
\centerline{$^{52}$University of Illinois at Chicago,                         
            Chicago, Illinois 60607, USA}                                     
\centerline{$^{53}$Northern Illinois University, DeKalb, Illinois 60115, USA} 
\centerline{$^{54}$Northwestern University, Evanston, Illinois 60208, USA}    
\centerline{$^{55}$Indiana University, Bloomington, Indiana 47405, USA}       
\centerline{$^{56}$University of Notre Dame, Notre Dame, Indiana 46556, USA}  
\centerline{$^{57}$Purdue University Calumet, Hammond, Indiana 46323, USA}    
\centerline{$^{58}$Iowa State University, Ames, Iowa 50011, USA}              
\centerline{$^{59}$University of Kansas, Lawrence, Kansas 66045, USA}         
\centerline{$^{60}$Kansas State University, Manhattan, Kansas 66506, USA}     
\centerline{$^{61}$Louisiana Tech University, Ruston, Louisiana 71272, USA}   
\centerline{$^{62}$University of Maryland, College Park, Maryland 20742, USA} 
\centerline{$^{63}$Boston University, Boston, Massachusetts 02215, USA}       
\centerline{$^{64}$Northeastern University, Boston, Massachusetts 02115, USA} 
\centerline{$^{65}$University of Michigan, Ann Arbor, Michigan 48109, USA}    
\centerline{$^{66}$Michigan State University,                                 
            East Lansing, Michigan 48824, USA}                                
\centerline{$^{67}$University of Mississippi,                                 
            University, Mississippi 38677, USA}                               
\centerline{$^{68}$University of Nebraska, Lincoln, Nebraska 68588, USA}      
\centerline{$^{69}$Princeton University, Princeton, New Jersey 08544, USA}    
\centerline{$^{70}$State University of New York, Buffalo, New York 14260, USA}
\centerline{$^{71}$Columbia University, New York, New York 10027, USA}        
\centerline{$^{72}$University of Rochester, Rochester, New York 14627, USA}   
\centerline{$^{73}$State University of New York,                              
            Stony Brook, New York 11794, USA}                                 
\centerline{$^{74}$Brookhaven National Laboratory, Upton, New York 11973, USA}
\centerline{$^{75}$Langston University, Langston, Oklahoma 73050, USA}        
\centerline{$^{76}$University of Oklahoma, Norman, Oklahoma 73019, USA}       
\centerline{$^{77}$Oklahoma State University, Stillwater, Oklahoma 74078, USA}
\centerline{$^{78}$Brown University, Providence, Rhode Island 02912, USA}     
\centerline{$^{79}$University of Texas, Arlington, Texas 76019, USA}          
\centerline{$^{80}$Southern Methodist University, Dallas, Texas 75275, USA}   
\centerline{$^{81}$Rice University, Houston, Texas 77005, USA}                
\centerline{$^{82}$University of Virginia, Charlottesville,                   
            Virginia 22901, USA}                                              
\centerline{$^{83}$University of Washington, Seattle, Washington 98195, USA}  
}                                                                             


\date{}

\begin{abstract}
We report a measurement of the $B^0_{s}$ lifetime in the 
semileptonic decay channel $B^0_{s}\rightarrow D^-_s \mu^{+}\nu X$ 
(and its charge conjugate), using approximately 0.4 fb$^{-1}$ of 
data collected with the D0 detector during 2002--2004. Using 5176 
reconstructed $D^-_s \mu^{+}$ signal events, we have 
measured the $B^0_s$ lifetime to be 
$\tau(B^0_{s}) = 1.398 \pm 0.044$~$(\mbox{stat}) ^{+0.028}_{-0.025}$
$(\mbox{syst})\ \mbox{ps}$.
This is the most precise measurement of the $B_s^0$ lifetime to date.
\end{abstract}

\maketitle 

Measurements of the lifetimes of different $b$ hadrons allow tests of 
the mechanism of heavy hadron decay.
The spectator model predicts that all hadrons with the same heavy flavor 
content have identical lifetimes. 
However, observed charm and bottom hadron lifetimes suggest that 
non-spectator effects, such as interference between contributing amplitudes,
are not negligible in heavy hadron decays. 
This implies that a mechanism beyond the simple spectator model is required.
An effective theory called the Heavy Quark Expansion (HQE)~\cite{bigi} 
includes such effects and predicts lifetime differences among 
the different bottom hadrons. In particular, a difference of the order of 1\%
is predicted between $B^0$ and $B_s^0$ mesons.
The measurement of the flavor-specific $B^0_s$ lifetime using semileptonic 
decays is also useful in determining the decay width difference between 
the light and heavy mass eigenstates of the $B^0_s$ meson, which is an 
equal mixture of $CP$ eigenstates that correspond to mass eigenstates in 
the absence of $CP$ violation in the $B^0_s$ system.

In this Letter, we present a high-statistics measurement of the $B^0_{s}$ 
lifetime, using a large sample of semileptonic $B_s^0$ decays collected 
in  $p\bar p$ collisions at $\sqrt{s} = 1.96$~TeV with the D0 detector at 
the Fermilab Tevatron Collider in 2002 -- 2004.
The data correspond to approximately 0.4 fb$^{-1}$ of integrated 
luminosity. $B_s^0$ mesons were identified through their semileptonic decay
$B_s^0 \rightarrow D_s^- \mu^+ \nu X$~\footnote{Unless otherwise stated, 
charge-conjugate states are implied.}, where the $D_s^-$ meson 
decays via $D_s^-\rightarrow \phi \pi^-$, followed by 
$\phi \rightarrow K^+ K^-$. 

The D0 detector is described in detail elsewhere~\cite{run2det}. The detector
components most important to this analysis are the central tracking and muon 
systems. The D0 central-tracking system consists of a silicon microstrip 
tracker (SMT) and a central fiber tracker (CFT), both located within a 2 T 
superconducting solenoidal magnet, with designs optimized for tracking and
vertexing at pseudorapidities $|\eta|<3$ and $|\eta|<2.5$, respectively (where
$\eta$ = $-$ln[tan($\theta$/2)]).
A liquid-argon and uranium calorimeter has a
central section covering pseudorapidities up to
$\approx 1.1$, and two end calorimeters that extend the coverage
to $|\eta|\approx 4.2$~\cite{run1det}. The muon system is located outside the 
calorimeters and has pseudorapidity coverage  $|\eta|<2$. 
It consists of a layer of tracking detectors and scintillation 
trigger counters in front of 1.8 T toroids, followed by two similar layers
after the toroids~\cite{run2muon}.

Events with semileptonic $B$-meson decays were selected 
using inclusive single-muon triggers in a three-level 
trigger system. The triggers used did not impose any impact parameter
criterion and were shown to not bias the 
lifetime measurement. 
Off-line, muons were identified by extrapolation of
the muon track segments, formed by the hits in the muon system,
to the tracks found in the central tracking system. 
Each muon was required to have a momentum $p>3$~GeV/$c$ 
and a transverse momentum $p_T>2$~GeV/$c$.

The primary vertex of each $p\bar{p}$ interaction was defined by all 
available well-reconstructed tracks~\cite{PVref} and constrained by 
the mean beam-spot position. The latter was 
updated every few hours.
The resolution of the reconstructed primary vertex was typically 20 $\mu$m
in the transverse plane and 40 $\mu$m in the 
beam direction.

To reconstruct $D_s^- \rightarrow \phi \pi^-$ decays,
tracks with $p_T > 1.0$ GeV/$c$ were assigned the kaon mass and 
oppositely charged pairs were combined  to 
form a $\phi$ candidate. Each $\phi$ candidate was required to have a mass in 
the range 1.008 -- 1.032 GeV/$c^2$, compatible with the reconstructed 
$\phi$ mass at D0. The $\phi$ candidate was then combined with another 
track of $p_T > 0.7$ GeV/$c$. For the ``right-sign'' combinations, we required 
the charge of the track to be opposite to that of the muon and assigned 
the pion mass to this track. All selected tracks
were required to have at least one SMT hit and one CFT hit.
The three tracks selected were combined to form a common vertex
(the $D_s^-$ vertex) with 
a confidence level 
greater than 0.1\%. The $D_s^-$ candidate was 
required to have $p_T>3.5$ GeV/$c$.

The secondary vertex, where the $B_s^0$ decays to a muon and a $D_s^-$ meson,
 was obtained by finding the intersection of the trajectory of the muon 
track and the flight path of the $D_s^-$ candidate. The confidence level
of that vertex had to be greater that 0.01\%.
The reconstructed $D_s^-$ decay 
vertex was required to be displaced from the primary vertex 
in the direction of the $D_s^-$ momentum. 

The helicity angle, $\Phi$, defined as the angle between the 
directions of the $K^-$ and $D_s^-$ in the $\phi$ rest frame, has a 
distribution proportional to $\cos^2 \Phi$. A cut of $|\cos \Phi| > $ 0.4 
was applied to further reduce combinatorial background, which was found to 
have a flat distribution. In order to suppress the physics
background originating from $D^{(*)}D^{(*)}$ processes~\footnote{$D^{(*)}$ 
denotes either $D$, $D^*$ or $D^{**}$.}, we required that the transverse
momentum of the muon with respect to the $D_s^-$ meson, $p_{Trel}$, exceed
 2 GeV/$c$. 
The $D_s^-\mu^+$ invariant mass was also restricted to $3.4 - 5.0$ GeV/$c^2$,
to be consistent with a 
$B$-meson candidate. Since the number of tracks near the $B_s^0$
candidate tends to be small, we required the isolation ${\cal I} = p^{tot}(\mu^+ D_s^-) / ( p^{tot}(\mu^+ D_s^-) + \sum p_i^{tot})>0.65$,  
where the sum
$\sum p_i^{tot}$ was taken over all charged particles in the cone 
$\sqrt{(\Delta\phi)^2+(\Delta\eta)^2} < 0.5$, with $\Delta\phi$ and 
$\Delta\eta$ being the azimuthal angle and the
pseudorapidity with respect to the $(\mu^+ D_s^-)$ direction.
The muon, kaon, and pion tracks were not included in the sum. 

The lifetime of the $B_s^0$, $\tau$, is related to the decay length in the 
transverse plane, $L_{xy}$,  by $L_{xy} = c\tau p_T/m$, where $p_T$ is the 
transverse momentum of the  $B_s^0$ and $m$ is its invariant mass. $L_{xy}$ 
is defined as the displacement of the $B_s^0$ vertex from the primary vertex 
projected onto the transverse momentum of the $D_s^-\mu^+$ system.
Since the $B_s^0$ meson is not fully reconstructed, $p_T(B_s^0)$ is
estimated by
$p_T(D_s^-\mu^+)/K$, where the correction factor, 
$K=p_T(D_s^-\mu^+)/p_T(B_s^0)$ is
determined using Monte Carlo (MC) methods.
The quantity used to extract the $B_s^0$ lifetime is called the 
pseudo-proper decay length (PPDL). The correction factor $K$ was 
applied statistically when extracting $c\tau(B_s^0)$ from the PPDL in 
the lifetime fit.

\begin{figure}
\includegraphics[scale=0.43]{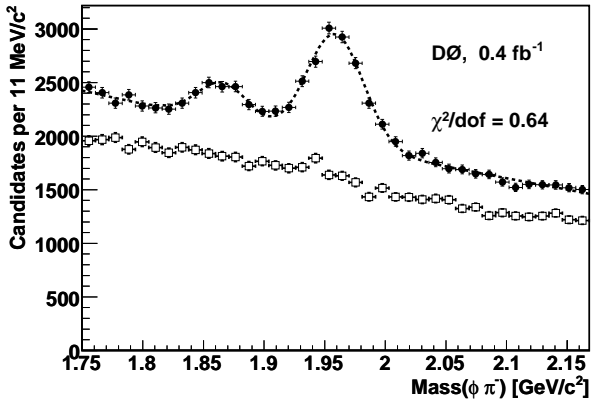}
\caption{\label{fig_1}The mass distribution of $\phi\pi^-$ candidates.
Points with er\-rors bars show the ``right-sign'' $D^-_s\mu^+$ combinations, 
and the open squares show the corresponding ``wrong-sign'' distribution.
The dashed curve represents the result of the fit to the ``right-sign'' 
combinations. The two peaks are associated with the $D^-$ and $D_s^-$ 
mesons, respectively.
}
\end{figure}

In the cases with more than one $B_s^0$ candidate per event, 
we chose the one with the highest vertex confidence level.
We also required the PPDL uncertainty to be less than 500 $\mu$m.
The resulting invariant mass distribution of the $D_s^-$ candidates is 
shown in Fig.~\ref{fig_1}. The distribution for 
``right-sign'' $D_s^-\mu^+$ candidates was fitted using a Gaussian, to 
describe the signal, and a second-order polynomial, 
to describe the combinatorial background. A second Gaussian was included
 for the Cabibbo-suppressed $D^-\rightarrow \phi \pi^-$ decay. 
The best fit result is shown in the same figure. 
The fit yields a signal of $5176 \pm 242$ (stat) $\pm\ 314$ (syst) 
$D_s^-$ candidates and a mass of $1958.8\pm 0.9$ MeV/$c^2$.
The width of the $D_s^-$ Gaussian is $22.6\pm 1.0$ MeV/$c^2$. 
The systematic uncertainty comes from the fit.
For the $D^-$ meson, the fit yields
$1551$ events. Figure~\ref{fig_1} also shows the invariant mass 
distribution of the ``wrong-sign'' candidates. The observed shift in the
$D^-_s$ mass is consistent with known issues associated with the
calibration of the D0 track momenta.  The contribution to the mass region
from reflected states was found to be negligible. 
Studies confirmed that this mass shift
introduces no significant residual bias in the lifetime determination.

MC samples were generated using {\sc Pythia}~\cite{pythia} for
the production and hadro\-niza\-tion phase, and {\sc EvtGen}~\cite{EvtGen} 
for decaying the $b$ and $c$ hadrons. Branching ratios from the
PDG have been used when available. Detector acceptance and smearing were 
taken into account using the
full D0 detector simulation based on {\sc geant}~\cite{Geant}.
Generated MC signal samples include contributions from 
$D_s^{-}     \mu^+ \nu$,
$D_s^{*-}    \mu^+ \nu$, 
$D_{s0}^{*-} \mu^+ \nu$, 
$D_{s1}^{'-} \mu^+ \nu$, and
$D_s^{(*)-} \tau^+ \nu$.

Apart from the background due to combinatorial processes such as a prompt muon 
and an identified $D_s^-$ meson, there are several real physics processes that
 produce a muon and a $D_s^-$ meson,  where neither comes from the semileptonic
 decay of the $B_s^0$ meson. 
These ``right-sign''  $D_s^-\mu^+$ combinations are included in the signal
sample and are defined as ``physics backgrounds.'' 
Prompt $D_s^-$ mesons from $c\bar{c}$ production at the interaction 
point can combine with high-$p_T$ muons generated either via direct 
production or in charm decays.
These $c\bar{c}$ background events are expected to have
very short lifetimes and thus could introduce a significant bias
in the $B_s^0$ lifetime measurement. 
Backgrounds
that originate from $\bar{B}$ mesons 
and provide the $D_s^-\mu^+$ final state, but not via the semileptonic decay 
$B_s^0\rightarrow D_s^-\mu^+\nu X$,
are called non-$B_s^0$ backgrounds. 
This kind of background is expected to have a relatively long lifetime, 
thus its effect on the $B_s^0$ lifetime fit is smaller than that of the 
charm background. 
There are three sources of such events:
$\bar{B}^0\rightarrow D_s^{(*)-}D^{(*)+}X$, $B^-\rightarrow D_s^{(*)-}\bar{D}^{(*)0}X$, and 
$\bar{B}_s^0\rightarrow D_s^{(*)-}D^{(*)}X$, 
where the charm meson accompanying the $D_s^{(*)-}$, which decays to 
$\phi\pi^-$, decays semileptonically.
The momentum of the muon coming
from the decay of the $D^{(*)}$ is softer than that for the signal,
since it comes from the decay of a secondary charm hadron.  This
implies that the contribution of these modes to the signal sample
is reduced by the kinematic cuts. We found the fractional contribution
of the backgrounds to the signal region to 
be $(10.0 \pm 7.0) $\% for $c\bar{c}$ background and 
$(11.3^{+5.3}_{-3.6})$\% for non-$B_s^0$ backgrounds.

The lifetime of the $B_s^0$ was found using a fit to the PPDL distribution.
We defined a signal sample using the 
$D_s^-$ mass distribution in the region from 1913.6 MeV/$c^2$ to 2004.0 
MeV/$c^2$, corresponding to $\pm2\sigma$ from the fitted mean mass. The PPDL distribution of the combinatorial background events
contained in the signal sample was defined using ``right-sign'' events from the 
$D_s^-$ sidebands (1755.3 -- 1800.5 MeV/$c^2$ and 2117.1 -- 2162.3 MeV/$c^2$) 
and ``wrong-sign'' events between 1755.3 and 2162.3 MeV/$c^2$.
The combinatorial background due to random track combinations was modeled by the
sideband sample events.
This assumption is supported by the mass distribution
of the ``wrong-sign'' combinations where no enhancement is visible in the 
$D_s^-$ mass region.

The PPDL distribution obtained from the signal sample was fitted
using an unbinned maximum log-likelihood method. Both the $B_s^0$ lifetime 
and the background shape were determined in a simultaneous fit to the signal 
and background samples.
The likelihood function ${\cal{L}}$ is given by

\begin{equation}
{\cal{L}} = {\cal{C}}_{sig} \prod_{i}^{N_S} [     f_{sig} {\cal{F}}^i_{sig} + 
                             (1 - f_{sig} ){\cal{F}}^i_{bck} ]
            \prod_{j}^{N_B} {\cal{F}}^j_{bck},
\end{equation}

\noindent where $N_S$, $N_B$ are the number of events in the signal and 
background samples and $f_{sig}$ is the ratio of $D_s^-$ 
signal events obtained from the $D_s^-$ mass distribution fit to the total 
number of events in the signal sample. To constrain $f_{sig}$, we factored 
in an additional likelihood term using the number of $D_s^-$ signal events 
observed from the invariant mass distribution, and its uncertainty,
 ${\cal{C}}_{sig}$. 

Since the current world average width difference between the light and
heavy mass eigenstates ($\Delta\Gamma_s$) of the $B_s^0$ system is small~\cite{hfag}
compared with the current precision of the data,
 we used for the signal probability distribution 
function (PDF), ${\cal F}^i_{sig}$, a normalized single exponential decay 
function convoluted with a Gaussian resolution function. 
The $K$-factor correction was also convoluted
with the exponential decay function.
Since a priori, we do not know the decay length uncertainty,
which we estimated on an event-by-event basis, an overall global scale 
factor, $s$, was introduced
as a free parameter in the $B_s^0$ lifetime fit.
The events from non-$B_s^0$ background were taken into account in the fit 
by including similar PDFs to those in the signal but using fixed parameters 
according to the world-average values~\cite{pdg}.
A different $K$-factor distribution was also
used for each process. To take into account the $c\bar{c}$
background, we used a Gaussian distribution with fixed parameters.
These contributions were evaluated and parametrized using MC
methods following a similar procedure as for the signal evaluation.

The combinatorial background sample, ${\cal F}^i_{bck}$, was para\-me\-tri\-zed 
us\-ing a Gaussian distribution function
for the resolution plus several exponential decays: two for the negative values in 
the PPDL distribution (one short and one long component) and two
for the positive values of the distribution.

\begin{figure}
\includegraphics[scale=0.43]{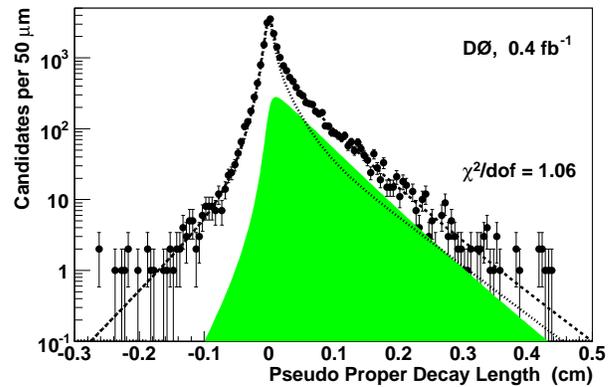}
\caption{\label{fig_2}Pseudo-proper decay length distribution for $D_s^-\mu^+$ candidates with
the result of the fit superimposed as the dashed curve. The dotted curve represents the combinatorial
background and the filled area represents the $B_s^0$ signal.}
\end{figure}

Figure~\ref{fig_2} shows the PPDL distribution of the 
$D_s^-\mu^+$ signal sample with the fit result superimposed (dashed curve). The 
dotted curve represents the sum of the background probability function over the events 
in the signal sample. The $B_s^0$ signal is represented by the filled area.

To test the resolutions, pulls, fitting, and selection criteria,
we performed detailed studies using MC samples and found no
significant bias in our analysis procedure.
In order to study the stability of the $B_s^0$ lifetime measurement, 
we split the data sample into two parts according to 
different kinematic and geometric parameters, compared the fitted
results, and found the lifetimes consistent within their uncertainties.
We also varied the selection criteria 
and mass fit ranges, and did not observe any significant
shifts. We performed an extensive study of our fitting procedure, looking
for any possible bias using MC ensembles with statistics of the size of
our dataset and distributions as those in data. These  
samples were fitted, and the mean and width of the distributions of 
extracted parameters were found to be consistent with the fits to data.
One final check of the procedure involved performing a similar lifetime fit to
a control sample defined by the Cabibbo-suppressed decay 
$D^- \rightarrow \phi \pi^-$,
(see Fig.~\ref{fig_1}).
We found that 89.1\% of the sample comes from $B^0\rightarrow D^-\mu^+ X$, and
the $B^0$ lifetime to be 
$1.541 \pm 0.093$ ps, 
where the uncertainty is statistical only.
This result is in good agreement with the world average $B^0$ lifetime~\cite{pdg,hfag}. 

We considered and evaluated various sources of systematic uncertainties.
The major contributions come from the determination of
the combinatorial background, the model for the resolution, and the physics background.
To determine the systematics due to the uncertainty on the combinatorial background,
we tested other assumptions on the background samples: we used just 
the events in the sidebands, just the events in the wrong-sign combinations, 
and removed either the right sideband or the left sideband samples. 
We also modified 
the definitions of those samples, changing the mass window sizes and positions.
The largest difference in $c\tau$ observed in these variations of background modeling
was 4.3~$\mu$m, which was taken to be the systematic uncertainty due to this source.
The effect of uncertainty in the resolution of the decay length was studied using
an alternative global scale factor, $s$. We repeated the
lifetime fit with fixed values of $s$ obtained from MC samples and from a 
different lifetime analysis~\cite{psiphi}. 
Using a variation of the resolution scale by a factor of two
beyond these bounds, we found a 3.7~$\mu$m variation in $c\tau$.
The uncertainty from the physics background was evaluated by varying the branching 
fractions
of the different processes as well as the shapes of the lifetime templates, as given by 
their known lifetime values~\cite{pdg}. The variations were within one standard deviation
in each case.
 Assuming no
correlation between them, we added the effects of all the variations in quadrature and 
found a total 
contribution of  $^{+2.9}_{-4.2}\ \mu$m. Using a similar procedure, 
we evaluated the uncertainty coming from the determination of the $c\bar{c}$ background 
and found a difference of  $^{+2.3}_{-0.8}\ \mu$m.

To evaluate the uncertainty associated with the $K$ factor 
determination, we modified the kinematics of the event using a different decay model,
a different $p_T$ spectrum for the $b$ quark, and a different $p_T$ spectrum for the 
muon. We also varied the amount of each component, according to their uncertainty, of the 
$B_s^0 \rightarrow D_s^- \mu^+ X$ signal. In each case, the $K$ factor
was re-evaluated and the fit repeated. We added all $K$ factor variation effects in 
quadrature 
and found a total uncertainty of  $^{+3.6}_{-2.1}\ \mu$m. 

There are two 
requirements in our selection method that could potentially change the final result by 
altering the shape of the PPDL distribution:
$p_{Trel} > 2$ GeV/$c$ and
the positive displacement from the primary vertex of the reconstructed $D_s^-$ decay vertex.
Using MC methods, we evaluated their effects by removing them one at a time. 
The largest variation observed was $^{+3.0}_{-0.3}$ $\mu$m, and the selection 
efficiency is flat as a function of proper decay time.
The effect of a possible misalignment of the SMT system was tested in 
Ref.~\cite{psiphi}. We repeated the study using MC signal samples and observed the same shift
of $c\tau$ = 2~$\mu$m, which was taken as a systematic uncertainty due to possible misalignment. 
The total systematic uncertainty from all of these sources added in
quadrature is $^{+8.4}_{-7.6}\ \mu$m.

In summary, using an integrated luminosity of approximately 0.4 fb$^{-1}$, 
we have 
measured the $B_{s}^0$ lifetime in the decay channel $D_s^-\mu^+\nu X$ to be
$\tau(B_s^0) = 1.398 \pm 0.044\ \mbox{(stat)} \ ^{+0.028}_{-0.025}\ \mbox{(syst)}\ \mbox{ps}$.
Note that this measurement takes $\Delta\Gamma_s$ equal to zero. 
The extraction of the average lifetime $\bar{\tau}_s$ for 
$\Delta\Gamma_s \neq 0$ is straight forward~\cite{hfag}.
The result is in good agreement with previous experiments as well as 
the current world 
average value for all flavor-specific decays, 
$\tau(B_{s}^0) = 1.442 \pm 0.066$\ ps~\cite{others,hfag}. 
Our $B_s^0$ lifetime measurement is the most precise to date and exceeds 
the precision of the current world average 
measurement $\tau(B_{s}^0)_{PDG} = 1.461 \pm 0.057$\ ps~\cite{pdg}, where semileptonic 
and hadronic decays were combined. This measurement is approximately $2.5\sigma$ away from
the $B^0$ lifetime, more than the 1\% predicted by HQE.


We thank the staffs at Fermilab and collaborating institutions, 
and acknowledge support from the 
DOE and NSF (USA);
CEA and CNRS/IN2P3 (France);
FASI, Rosatom and RFBR (Russia);
CAPES, CNPq, FAPERJ, FAPESP and FUNDUNESP (Brazil);
DAE and DST (India);
Colciencias (Colombia);
CONACyT (Mexico);
KRF and KOSEF (Korea);
CONICET and UBACyT (Argentina);
FOM (The Netherlands);
PPARC (United Kingdom);
MSMT (Czech Republic);
CRC Program, CFI, NSERC and WestGrid Project (Canada);
BMBF and DFG (Germany);
SFI (Ireland);
The Swedish Research Council (Sweden);
Research Corporation;
Alexander von Humboldt Foundation;
and the Marie Curie Program.


\begin{thebibliography}{99}


\bibitem[*]{kurca}
On leave from IEP SAS Kosice, Slovakia.
\bibitem[\dag]{voutilainen}
Visitor from Helsinki Institute of Physics, Helsinki, Finland.
%

  \bibitem{bigi} I.I. Bigi {\it et al.}, in ``$B$ decays,'' 2nd edition, 
                  edited by 
                 S. Stone (World Scientific, Singapore, 1994).

  \bibitem{run2det} V. Abazov {\it et al.}, 
                    Nucl.\ Instrum.\ Meth.\ A {\bf 565}, 463 (2006).

  \bibitem{run1det} S. Abachi {\it et al.}, Nucl. Instrum. 
                    Methods Phys. Res. A {\bf 338}, 185 (1994).

  \bibitem{run2muon} V. Abazov {\it et al.}, Nucl. Instrum. 
                    Methods Phys. Res. A {\bf 552}, 372 (2005).

  \bibitem{PVref} J. Abdallah {\it et al.},
                  Eur. Phys. J. C {\bf 32}, 185, (2004).

  \bibitem{pythia} T. Sj\"{o}strand {\sl et al.}, Comp. Phys. Commun. {\bf 135}, 238 (2001), v6.2.

  \bibitem{EvtGen} D. J. Lange, Nucl. Instrum. Methods Phys. Res. A {\bf 462}, 152 (2001), v00-11-07.

  \bibitem{Geant} R. Brun {\sl et al.}, CERN Report No. DD/EE/84--1, 1984.

  \bibitem{pdg} Particle Data Group,
                S. Eidelman {\sl et al.}, Phys. Lett. B {\bf 592}, 1 (2004).

  \bibitem{hfag} 
    K. Anikeev {\it et al.} (Heavy Flavor Averaging Group), hep-ex/0505100.

  \bibitem{psiphi} V. Abazov {\it et al.}, Phys. Rev. Lett.
 {\bf 94}, 042001 (2005). 

  \bibitem{others} D. Buskulic {\sl et al.}, Phys. Lett. B {\bf 377}, 205 (1996);
                   K. Ackerstaff {\sl et al.}, Phys. Lett. B {\bf 426}, 161 (1998);
                   F. Abe {\sl et al.}, Phys. Rev. D {\bf 59}, 032004 (1999);
                   P. Abreu {\sl et al.}, Eur. Phys. J. C {\bf 16}, 555 (2000).

\end{thebibliography}
\end{document}